\documentstyle[epsfig,aps,twocolumn]{revtex}
\title{Relation between complementarity and nonlocality under
decoherence}
\author{R. Filip\footnote{email:filip@optnw.upol.cz, tel:+420-68-5631572,
fax:+420-68-5224246}\\
Department of Optics, Palack\' y University,\\
17. listopadu 50,  772~07 Olomouc, \\ Czech Republic}
\date{\today}
\begin{document}
\maketitle
\begin{abstract}
Relations connecting violation of any Bell inequalities and
the complementarity between visibility and distinguishability
in the interferometric experiments with
different sources of decoherence are presented.
A boundary of local-realistic explanation of the which-way
complementarity is discussed in dependence on choice of
independent or collective tests of nonlocality.
\end{abstract}
PACS number(s): 42.50.Dv\newline

\section{Introduction}

Quantum-mechanical nonlocality
\cite{Einstein35,Bohr35,Bohm52,Bell64,Clauser69}
and complementarity between which-way
information and visibility \cite{Greenberger88,Jaeger93,Jaeger95,Englert96a}
are well-known manifestations of
the quantum-mechanical microworld.
However, a nature of both the effects is identical: it is
quantum entanglement shared
between parts of total system.
In the consequence, a natural question arises:
what is a relation between complementarity and nonlocality in
commonly used interferometric experiments?
Consider a fact that there exist the states
which are entangled however do not violate any Bell-type inequalities
\cite{Werner89,Popescu94,Horodecki95} and that relation between
visibility and distinguishability can be also assisted by classical
correlations, it seems to be important to find a
boundary for explanation of complementarity experiments in terms of
local-realistic theories (LRT).
Up to this boundary, no LRT cannot explain the complementary results of
experiment, whereas below this boundary some LRT can exist.
In addition, it is well known that multiple
copies of entangled states, which do not violate Bell inequalities,
could be distilled by the local operations and classical
communications to produce a state violating the Bell
inequality \cite{Gisin96,Horodecki96}. Arising question is:
how changes a connection between complementarity and nonlocality using
the collective tests of nonlocality ?

For the simplicity, we consider both the analysed effects
based on the measurements of a bipartite system 
compounded from qubit $A$ and $B$ systems .
It will be assumed that the measurements are performed
by two different experimentalists (Alice and Bob) sharing
state which arises from Bob's monitoring of Alice's system.
Fortunately, for two-qubit systems the questions concerning to violating
of any Bell inequalities \cite{Horodecki95},
entanglement distillation \cite{Bennett96,Horodecki97} and
inseparability of mixed state \cite{Peres96} have been solved completely.
In addition, it was proved \cite{Zukowski01,Brukner01}
that necessary and sufficient condition for the local realistic
description of two-qubit correlations is equivalent to condition on
violating of any Bell inequality \cite{Horodecki95}. Due to these
important facts, a sharp boundary between LRT and quantum
mechanical explanation of which-way complementarity could be proposed.
In this paper, it is shown, on simply example of interferometric
experiments, how nonlocality, entanglement and quantum information
features are related to the complementarity under
influence of different kind of decoherence in the system and
meter. The most attractive are cases, where
the complementarity effects could be
explained only by quantum mechanics and not also some kind of local
realistic theory. The individual and collective tests of nonlocality are
discussed in this consequence.

\section{Decoherence-free case}

To begin analysis, a decoherence-free case of interferometric
experiment is shortly examined.
At the input, the qubit $A$ is prepared in the state
\begin{equation}
|\Psi_{A}\rangle=\sqrt{r}|\uparrow\rangle-\sqrt{1-r}|\downarrow\rangle
\end{equation}
which can be treated as the superposition of a possible ``paths'' in
interferometer.
The qubit $B$ is considered in the state $|\downarrow\rangle$.
Commonly used nondemolition monitoring of the qubit $A$ by the qubit $B$
is considered to acquire an information about the individuality of the
state in superposition
\begin{eqnarray}\label{nondemo}
|\uparrow\rangle_{A}|\downarrow\rangle_{B} &\longrightarrow &
|\uparrow\rangle_{A}|\downarrow\rangle_{B},\nonumber\\
|\downarrow\rangle_{A}|\downarrow\rangle_{B} &\longrightarrow &
\sqrt{1-D^2}|\downarrow\rangle_{A}
|\downarrow\rangle_{B}+D|\downarrow\rangle_{A}
|\uparrow\rangle_{B},
\end{eqnarray}
where $D$ is a measure of distinguishability due to nondemolition
monitoring.
Here, the distinguishability is independent on a
possible measurement performed by Bob; it is considered rather as parameter of
interaction between systems $A$ and $B$.
After monitoring, an output state is given in the following form
\begin{eqnarray}\label{state1}
|\Psi\rangle=\sqrt{r}|\uparrow\rangle_{A}|\downarrow\rangle_{B}-
\sqrt{1-r}(\sqrt{1-D^2}|\downarrow\rangle_{A}|\downarrow\rangle_{B}+\nonumber\\
+D|\downarrow\rangle_{A}|\uparrow\rangle_{B})
\end{eqnarray}
and if one performs phase shift operation
\begin{equation}\label{shift}
|\uparrow\rangle_{A}\longrightarrow
\exp(-i\phi)|\uparrow\rangle_{A},\quad
|\downarrow\rangle_{A}\longrightarrow |\downarrow\rangle_{A},
\end{equation}
and specific rotations on the
qubit $A$
\begin{eqnarray}\label{rotat}
|\uparrow\rangle_{A} &\longrightarrow &
\frac{1}{\sqrt{2}}(|\uparrow\rangle_{A}+|\downarrow\rangle_{A}),\nonumber\\
|\downarrow\rangle_{A} &\longrightarrow &
\frac{1}{\sqrt{2}}(-|\uparrow\rangle_{A}+|\downarrow\rangle_{A}),
\end{eqnarray}
then the visibility
\begin{equation}\label{visib}
V=\frac{p_{\mbox{max}}(\phi)-p_{\mbox{min}}(\phi)}
{p_{\mbox{max}}(\phi)+p_{\mbox{min}}(\phi)}
\end{equation}
of interference measured on the qubit $A$ is related
to defined distinguishability $D$ and the predictability
$P=|p_{\downarrow}-p_{\uparrow}|=\sqrt{|1-2r|}$ in the following
way
\begin{equation}\label{comple1}
\frac{V^{2}}{1-P^{2}}+D^{2}=1.
\end{equation}
Using the overlap $O=\sqrt{1-D^{2}}$ and the unpredictability
$U=\sqrt{1-P^{2}}$, the relation can be re-arrangered to
\begin{equation}\label{vou}
V=OU,
\end{equation}
where $V,D,O,U\in\langle 0,1 \rangle$.
The predictability $P$ (and inversely unpredictability $U$) describes a priori
information about ``path'' in the interferometer, whereas
distinguishability $D$ (and overlap $O$) are rather connected with
efficiency of path monitoring by the meter system $B$.
By a decreasing of unpredictability in system $A$ and overlap in system
$B$, the visibility of interference vanishes as can be seen in
Eq.~(\ref{vou}). For given predictability $P$, the relation
(\ref{comple1}) can be interpreted as a particular relation of complementarity
between which-way information expressed by distinguishability $D$
and visibility $V$.

Recently Gisin \cite{Gisin91,Popescu92} showed that any entangled pure
state of a pair of qubits violates the Clauser-Horne-Shimony-Holt (CHSH)
form of Bell inequalities. Particularly,
in considered case of pure state (\ref{state1}), the dependence of maximal
any Bell-CHSH inequality violation
\footnote{The Horodecki theorem states that a two qubits state
$\hat{\rho}$ violates a Bell-CHSH inequality if and only if
$M(\rho)>1$, where $M$ is real-valued function of the density matrix
$\hat{\rho}$. To define $M$, one needs the $3\times3$ matrix $T$ with
entries $T_{ij}=\mbox{Tr}\{\rho(\sigma_{i}\otimes\sigma_{j})\}$. Then,
$M(\rho)$ the sum of the two largest eigenvalues of the Hermitian
matrix $\hat{T}^{\dag}\hat{T}$. A maximal value of Bell-CHSH factor
$B_{\mbox{BSHS}}=C(a,b)+C(a,b')+C(a',b)-C(a',b')$ is given in the
following form $B_{\mbox{max}}=2\sqrt{M(\rho)}$.}
on the distinguishability $D$ and
predictability $P$ can be determined
\begin{equation}\label{Bell1}
B_{\mbox{max}}=2\sqrt{1+D^{2}U^{2}}
\end{equation}
and is depicted in Fig.~1.
In this decoherence-free case, for every $D\not= 0$ and $U\not= 0$
the state shared between Alice and Bob used to test intrinsic quantum
complementarity, because it violates Bell-CHSH inequality.
The complementarity is completely
assisted by the quantum-mechanical nonlocality. No local-realistic
theory can explain these results.
On the other hand, for this pure state case,
we assume two different kinds of decoherence in the system and meter.
These cases will proceed separately and general results are presented
in the conclusion.

\section{System decoherence}

First, we consider an additional observer (Eve), which nondemolitionally
monitored the system $A$ in the following way
\begin{eqnarray}\label{moni}
|\uparrow\rangle_{A}|\downarrow\rangle_{E} &\longrightarrow &
|\uparrow\rangle_{A}|\downarrow\rangle_{E},\nonumber\\
|\downarrow\rangle_{A}|\downarrow\rangle_{E} &\longrightarrow &
(R|\downarrow\rangle_{A}
|\downarrow\rangle_{E}+\sqrt{1-R^{2}}|\downarrow\rangle_{A}
|\uparrow\rangle_{E}),
\end{eqnarray}
where $R$ is a measure of robustness of Alice state against the
decoherence due to Eve's intervence. Her action leads to the following state
\begin{eqnarray}
\frac{1}{\sqrt{2}}(|\uparrow\rangle_{A}
|\downarrow\rangle_{B}|\downarrow\rangle_{E}-|\downarrow\rangle_{A}
(\sqrt{1-D^2}|\downarrow\rangle_{B}+\nonumber\\
+D|\uparrow\rangle_{B})
(R|\downarrow\rangle_{E}+\sqrt{1-R^{2}}|\uparrow\rangle_{E})).
\end{eqnarray}
In this case, if one performs a phase shift (\ref{shift})
and the specific rotations (\ref{rotat}) on the
qubit $A$,
then the visibility $V$ in (\ref{visib})
of interference measured on the qubit $A$ is related
to the distinguishability $D$ and the robustness $R$ in the following
way
\begin{equation}\label{comple2}
\frac{V^{2}}{R^{2}}+D^{2}=1.
\end{equation}
It is similar to formulae (\ref{comple1}), if the robustness $R$ is
identificated with unpredictability $U$, however the action is
rather nonunitary here.
Using the defined visibility $V_{0}=\sqrt{1-D^{2}}$ without decoherence for the
same distinguishability,
the robustness $R$ can be evaluated in the following form
\begin{equation}
R=\frac{V}{V_{0}}.
\end{equation}
Thus, the robustness $R$ can be also interpreted as ratio between
visibility with decoherence $V$ and decoherence-free visibility $V_{0}$.

However, irrespective to formal analogue between (\ref{comple1}) and
(\ref{comple2}), a question is still opened: is it possible to explain the
complementarity by a hidden local-variable theory, i.e. can the
complementary observations be predetermined before the measurement
by an existing local element of physical reality.
One possibility is to test the Bell inequalities on the state shared by
Alice and Bob. To test it, a general CHSH form of the Bell-inequalities
\cite{Horodecki95}
can be used for general state of $\mbox{2}\times\mbox{2}$ system.
After calculations, it can be found that maximal value of the
Bell parameter $B_{\mbox{max}}$ is
\begin{equation}\label{Bell2}
B_{\mbox{max}}=2\sqrt{R^{2}+D^{2}},
\end{equation}
where dependence of $B_{\mbox{max}}$ on robustness $R$ and distinguishability $D$ is
depicted in Fig.~2. Contrary to an analogue between the complementarity
relations (\ref{comple1}) and (\ref{comple2}), Bell-CHSH inequality
violation differs substantially in these cases.
To overcome the maximal local-realistic correlations $B_{\mbox{max}}=2$,
the robustness $R$ must be large than the overlap
$O=\sqrt{1-D^{2}}$ between Bob's states
\begin{equation}
R>O
\end{equation}
or equivalently, the relation
\begin{equation}
D^{2}+R^{2}>1
\end{equation}
must be satisfied.
Particularly, for decoherence-free case $(R=1)$, complete relation
between visibility and distinguishability (\ref{comple2}) can be
explained only by the quantum mechanics and no local-realistic theory
can be constructed. With decreasing $R$,
using Eq.(\ref{comple2}), it can be found that for
the visibility $V$ it is sufficient to satisfy the following condition
\begin{equation}\label{cond}
V>1-D^2=O^2
\end{equation}
which cannot be explained by a local-realistic theory.
On the other hand, there are the regions of complementarity which can be
simulated by the local-realistic theory and are larger as $R$ decreases,
as is illustrated in Fig.~3. The inverse condition $V\le 1-D^2$ is a
necessary and sufficient condition for a state to be described in
terms of a local, hidden variable theory.

On the other hand, one can assume a generalized notion of nonlocality,
based on the distillation procedure and classical communications.
Generally, any inseparable mixed
state of a two-qubit systems can be distilled \cite{Bennett96}
to a singlet form by using
local operations and classical communications \cite{Horodecki97}.
Thus any nonseparable two-qubit system reveals
nonlocal properties if the sequential measurements are considered
collectively. Any inseparable mixed two-qubit state is nonlocal in this sense.
However, the collective tests of nonlocality describe a different
situation -- one can argue that if collective measurement on $n$ copies
of state $\hat{\rho}$ is needed to reveal nonlocality,
then it is the state $\hat{\rho}^{\otimes n}$ that is nonlocal rather
$\hat{\rho}$.

For two qubits, inseparability of state shared by Alice and Bob
can be theoretically tested by the
positive partial transposition criterion \cite{Peres96,Horodecki96}
\footnote{It has been shown that a state $\hat{\rho}$ of a $2\times 2$
system is inseparable if, and only if, its partial transposition
$\hat{\rho}^{T_{2}}$ is non-negative operator, i.e. if all eigenvalues
are nonnegative. Here the partial transposition $\hat{\rho}^{T_{2}}$
associated with the arbitrary product orthonormal
$|m\rangle_{A}|n\rangle_{B}$ basis is defined by the matrix elements in
this basis:
\begin{equation}
\rho^{T_{2}}_{mn,m'n'}=_{A}\langle m|_{B}\langle n|\hat{\rho}^{T_{2}}|n'
\rangle_{B}|m'\rangle_{A}=\rho_{mn',m'n}.
\end{equation}
Clearly, the matrix $\rho^{T_{2}}$ depends on the basis, but its
eigenvalues do not. Hence one can check separability using an arbitrary
product orthonormal basis in Hilbert space of two qubit system.}.
After tracing out Eve's states and some calculations,
a partially transposed density
matrix has three positive eigenvalues and one is negative for
$D\not=0$ or $R\not= 0$.
Thus except a total decoherence ($R=0$) or no distinguishability
($D=0$), Alice and Bob share an inseparable state which can be
distilled to maximal entangled state violating any Bell-CHSH inequality.
Even an imperfect complementarity with Eve's
intervence cannot be explained by the
hidden local-variable theory, if the distillation procedures are assumed.
This ``hidden nonlocality'' changes the boundary
of local-realistic explanation of the complementarity.

To quantificate an information acquired by Alice's monitoring, the
mutual information $I_{AB}=S_{A}+S_{B}-S_{AB}$, where $S_{A}$,
$S_{B}$ and $S_{AB}$ are von Neumann entropies of particular Alice's and
Bob's subsystems and total system, can be used
\cite{vonNeumann63,Wehrl78,BarnettPhoenix89}. After trace out the
Eve's states, the mutual information can be evaluated by the particular
entropies:
\begin{eqnarray}
S_{AB}&=&-\frac{1+R}{2}\ln\left(\frac{1+R}{2}\right)-
\frac{1-R}{2}\ln\left(\frac{1-R}{2}\right),\nonumber\\
S_{A}&=&-\frac{1+R\sqrt{1-D^{2}}}{2}\ln\left(\frac{1+R\sqrt{1-D^{2}}}{2}\right)-\nonumber\\
& &-\frac{1-R\sqrt{1-D^{2}}}{2}\ln\left(\frac{1-R\sqrt{1-D^{2}}}{2}\right),\nonumber\\
S_{B}&=&-\frac{1+\sqrt{1-D^{2}}}{2}\ln\left(\frac{1+\sqrt{1-D^{2}}}{2}\right)-\nonumber\\
& &-\frac{1-\sqrt{1-D^{2}}}{2}\ln\left(\frac{1-\sqrt{1-D^{2}}}{2}\right)
\end{eqnarray}
and is depicted in Fig.~4. For $R=1$, the mutual information varies
in interval $\langle 0,2\ln 2\rangle$ in dependence on $D$, whereas for $R=0$ in interval
$\langle 0,\ln 2\rangle$.

Without decoherence and for $D=1$, Bob
shared with Alice an additional ``quantum'' ebit,
in a comparison with the maximal decoherence
case. This additional ebit
can be used in quantum ``erasure'' procedure and
revealing of the visibility: one bit of information is damaged by
Bob's erasing
procedure and the other is send to Alice to distinguish between ``fringes''
and ``antifringes'' in the interference effect. Of course, for $R=0$ one
bit of information is on Eve's side and thus by performing same
procedure by Bob, the visibility cannot be revealed.
The violation of any Bell-CHSH inequalities can be simply expressed in
terms of mutual information: the acquired information is sufficient to
nonlocality exhibition if $I_{AB}$ is larger than a boundary value
$\bar{I}$
\begin{equation}
\bar{I}=-\frac{1+R^{2}}{2}\ln\left(\frac{1+R^{2}}{2}\right)-
\frac{1-R^{2}}{2}\ln\left(\frac{1-R^{2}}{2}\right).
\end{equation}
For given robustness $R$, the distinguishability must achieve some
threshold value in order to the Alice's information was sufficient to
exhibit the nonlocality, as can be seen in Fig.~4.
For almost unit robustness, only a small Bob's acquired information
is need to assist a nonlocal relation between Alice and Bob.
On the other hand, for strong decoherence (small robustness),
the nonlocal character of
mutual information can be considered if the distinguishability is
almost perfect.

\section{Meter decoherence}

In opposite to previous case with a
decoherence in the system, one can consider that
Eve monitors the meter in analogous way to (\ref{moni}).

Then the state arising in the total system can be considered
in the following way
\begin{eqnarray}
\frac{1}{\sqrt{2}}\left(|\uparrow\rangle_{A}
|\downarrow\rangle_{B}|\downarrow\rangle_{E}-|\downarrow\rangle_{A}
(\sqrt{1-D^2}|\downarrow\rangle_{B}|\downarrow\rangle_{E}+\right.\nonumber\\
\left.+D|\uparrow\rangle_{B}
\times(R|\downarrow\rangle_{E}+\sqrt{1-R^{2}}|\uparrow\rangle_{E}))\right),
\end{eqnarray}
where $R$ is a measure of Bob system robustness against Eve's
decoherence. For $R=1$ an information about path is extracted by means
of shared quantum entanglement between Alice and Bob,
whereas for $R=0$ only classical correlations are used to monitoring
procedure. After action of a phase shift (\ref{shift})
and the specific rotations (\ref{rotat}) on the
qubit $A$, the visibility $V$ in (\ref{visib})
of interference is related only to the distinguishability $D$
\begin{equation}\label{comple3}
V^{2}+D^{2}=1
\end{equation}
and is completely independent of the robustness $R$ of the meter.
The complementarity relation (\ref{comple3}) is completely identical to
decoherence-free one (\ref{comple1}) with $P=0$.
Thus based on this relation, one is not able to distinguishing between
classical and inherent quantum monitoring of the path.
On the other hand, the maximum of Bell-CHSH factor depends significantly
on the robustness
\begin{equation}
B_{\mbox{max}}=2\sqrt{(1-R^{2})(1-D^{2})^{2}+R^{2}+D^{2}},
\end{equation}
as it can be seen in Fig.~5.
Even if the complementarity relation is same as for idealized case
without decoherence, any Bell-CHSH inequality are violated
for $D\not= 0$ and $R\not= 1$ if
\begin{equation}\label{bound3}
1-D^{2}<\frac{R^{2}}{1-R^{2}}
\end{equation}
and for $R=1$ for every $D\not= 0$. For $R\geq\frac{1}{\sqrt{2}}$, the
violating Bell-CHSH inequality occurs for every $D\not= 0$, whereas for
$R<\frac{1}{\sqrt{2}}$, a boundary
is given by (\ref{bound3}).
There is a different character of the
boundary in comparison with previous system decoherence.
Under this boundary, the complementarity can be
assisted by a state non-violating any
Bell-CHSH inequality. Especially, if the complementarity is assisted by
classical correlations ($R=0$), no Bell-CHSH inequality violation occurs.
Thus, the results of complementarity experiment can be simulated by
local-realistic theory.
Irrespective to this, all the states are entangled
for $D\not=0$ and $R\not=0$ as can be proved using the
PPT criterion of separability.
Thus it contains a ``hidden
nonlocality'' in sense of collective tests of nonlocality, as it pointed
above. To compare an amount of Bob's acquired information with previous case,
the mutual information $I_{AB}=S_{A}+S_{B}-S_{AB}$ can be determined
from the particular entropies:
\begin{eqnarray}
S_{AB}&=&-\left(\frac{1}{2}+\frac{1}{2}\sqrt{1-D^{2}(2-D^{2})(1-R^{2})}\right)\times\nonumber\\
& &\times\ln\left(\frac{1}{2}+\frac{1}{2}\sqrt{1-D^{2}(2-D^{2})(1-R^{2})}\right)-\nonumber\\
& &-\left(\frac{1}{2}-\frac{1}{2}\sqrt{1-D^{2}(2-D^{2})(1-R^{2})}\right)\times\nonumber\\
& &\times\ln\left(\frac{1}{2}-\frac{1}{2}\sqrt{1-D^{2}(2-D^{2})(1-R^{2})}\right),\nonumber\\
S_{A}&=&-\frac{1+\sqrt{1-D^{2}}}{2}\ln\left(\frac{1+\sqrt{1-D^{2}}}{2}\right)-\nonumber\\
& &-\frac{1-\sqrt{1-D^{2}}}{2}\ln\left(\frac{1-\sqrt{1-D^{2}}}{2}\right),\nonumber\\
S_{B}&=&-\left(\frac{1}{2}+\frac{1}{2}\sqrt{(1-D^{2})^{2}(1-R^{2})}\right)\times\nonumber\\
& &\times\ln\left(\frac{1}{2}+\frac{1}{2}\sqrt{(1-D^{2})^{2}(1-R^{2})}\right)-\nonumber\\
& &-\left(\frac{1}{2}-\frac{1}{2}\sqrt{(1-D^{2})^{2}(1-R^{2})}\right)\times\nonumber\\
& &\times\ln\left(\frac{1}{2}-\frac{1}{2}\sqrt{(1-D^{2})^{2}(1-R^{2})}\right),\nonumber\\
\end{eqnarray}
and their dependence on the parameters $D$ and $R$ is depicted in Fig.~6.
From the boundary (\ref{bound3}), the condition on the violation of
Bell-CHSH inequality can be found in terms of information in following
form
\begin{eqnarray}
I>\bar{I}&=&
\left(\frac{1}{2}+\frac{1}{2}\frac{\sqrt{2}R^{2}}{\sqrt{1-R^{2}}}\right)
\ln\left(\frac{1}{2}+\frac{1}{2}\frac{\sqrt{2}R^{2}}{\sqrt{1-R^{2}}}\right)-\nonumber\\
& &-\left(\frac{1}{2}-\frac{1}{2}\frac{\sqrt{2}R^{2}}{\sqrt{1-R^{2}}}\right)
\ln\left(\frac{1}{2}-\frac{1}{2}\frac{\sqrt{2}R^{2}}{\sqrt{1-R^{2}}}\right).
\end{eqnarray}
From Fig.~4 and Fig.~6, it can be found that in terms of information the
Bell-CHSH inequality violation boundary has seemingly similar character, however the
system decoherence generates an information boundary for
any $R$, contrary to the meter decoherence where the information
boundary occurs only for $R<\frac{1}{\sqrt{2}}$.

The distinction between entanglement and classical correlations, used to
monitoring, seems to be crucial for the understanding of quantum
complementarity nature. Due to entanglement nature, the intrinsic
quantum complementarity experiments
has a ``reversible'' character: an information obtained by
monitoring can be erased in proper way to restore the visibility of
interference by postselection.
Similar procedure cannot be performed in classical case
and the complementarity experiments has an irreversible character.
The reason is that an
entanglement shared between Alice and Bob cannot be understand
in a classical realistic sense.

\section{Summary and conclusion}

A complementarity between measurement outputs need not be inherently
quantum mechanical and can be given by a local hidden-variable theory:
meter monitors the system in the classical sense of ``measuring'':
finding out what the state is.
On the other hand, if one assumes that the
complementarity is a pure quantum mechanical
feature and cannot be explained by the some realistic hidden-variable theory,
the mutual relation between distinguishability and interference must
be treated carefully under the influence of decoherence.

Summarizing both the decoherence cases, we assume the
system and meter decoherence simultaneously. The decoherence effects in
the system and meter are described by the system $R_{S}$ and
meter $R_{M}$ robustness. Then the visibility only depends on system
robustness $R_{S}$
\begin{equation}
\frac{V^{2}}{R_{S}^{2}}+D^{2}=1
\end{equation}
and maximal violation of any Bell-CHSH inequalities is expressed in
following form
\begin{equation}
B_{\mbox{max}}=2\sqrt{D^{2}(1-R_{M}^{2})(D^{2}-R_{S}^{2})+D^{2}R_{M}^{2}+
R_{S}^{2}}.
\end{equation}
Any Bell-CHSH inequalities are violated only if distinguishability $D$
satisfies the condition
\begin{eqnarray}\label{Dboun}
D^{2}&>&\frac{\alpha}{2}+\sqrt{(\alpha/2)^{2}+\beta},\nonumber\\
\alpha&=&R_{S}^{2}-\frac{R^{2}_{M}}{1-R^{2}_{M}},\nonumber\\
\beta&=&\frac{1-R_{S}^{2}}{1-R_{M}^{2}}.
\end{eqnarray}
The boundary on distinguishability (\ref{Dboun}) is depicted in Fig.~7
in dependence on meter and system robustness.
Then, one can be
sure that the complementary results of measurements
cannot be predetermined by an element of physical
reality and the quantum mechanics must be used to their correct
description. These conditions are
another expression of the quantum-mechanical nonlocality (in sense of the
Bell-inequality violation) in the terms of
complementary variables for the imperfect interferometric experiments.

In addition, the quantum complementarity is differently related to a
nonlocality, if the measurements are performed separately on each qubit
pair or collectively on several qubit pairs at once.
Separate imperfect measurements can lead to some nontrivial
bound on the violation of any Bell-CHSH inequalities for the
system-meter state. Apart from this, question about their local-realism
remain open. They do not violate this inequality but might violate some
other inequality.
On the other hand, in all the
considered cases the system-meter state can be distilled to pure
maximal entangled state and thus the nonlocality can be revealed even
for the imperfect complementarity experiments.

\medskip
\noindent {\bf Acknowledgments}
The author would like to thank Prof. J. Pe\v rina and L. Mi\v sta Jr.
for stimulating discussions about the quantum complementarity
and J. Fiur\' a\v sek for the reading of paper.
The work was supported by the project LN00A015 and CEZ:J14/98
of the Ministry of Education of Czech Republic.


\newpage

\begin{figure}
\caption{Decoherence-free case:
maximal violation of the CHSH-Bell inequalities in dependence
on distinguishability $D$ and unpredicability $U$.}
\end{figure}

\begin{figure}
\caption{System decoherence: maximal violation of Bell-CHSH inequalities in dependence on the
overlap $O$ and the robustness $R$.}
\end{figure}

\begin{figure}
\caption{System decoherence: visibility of interference $V$
in dependence on the distinguishability $D$ and the robustness $R$;
The thick-line establishes a boundary of
an explanation of the complementarity by the
local-realistic theory.}
\end{figure}

\begin{figure}
\caption{System decoherence: mutual information between Alice and Bob under
influence of Eve's system decoherence in dependence on the
distinguishability $D$ and robustness $R$; thick line is boundary for
a violation of Bell-CHSH inequalities.}
\end{figure}

\begin{figure}
\caption{Meter decoherence:
maximal violation of Bell-CHSH inequalities in dependence on
the robustness $R$ and distinguishability $D$.}
\end{figure}

\begin{figure}
\caption{Meter decoherence: mutual information between Alice and Bob under
influence of Eve's meter decoherence in dependence on the
distinguishability $D$ and robustness $R$; thick line is boundary for
a violation of Bell-CHSH inequalities.}
\end{figure}

\begin{figure}
\caption{Meter and system decoherence: lower bound of distinguishability
$D$ leading to violating of any
Bell-CHSH inequality in dependence on the system and meter
robustness $R_{S}$, $R_{M}$.}
\end{figure}

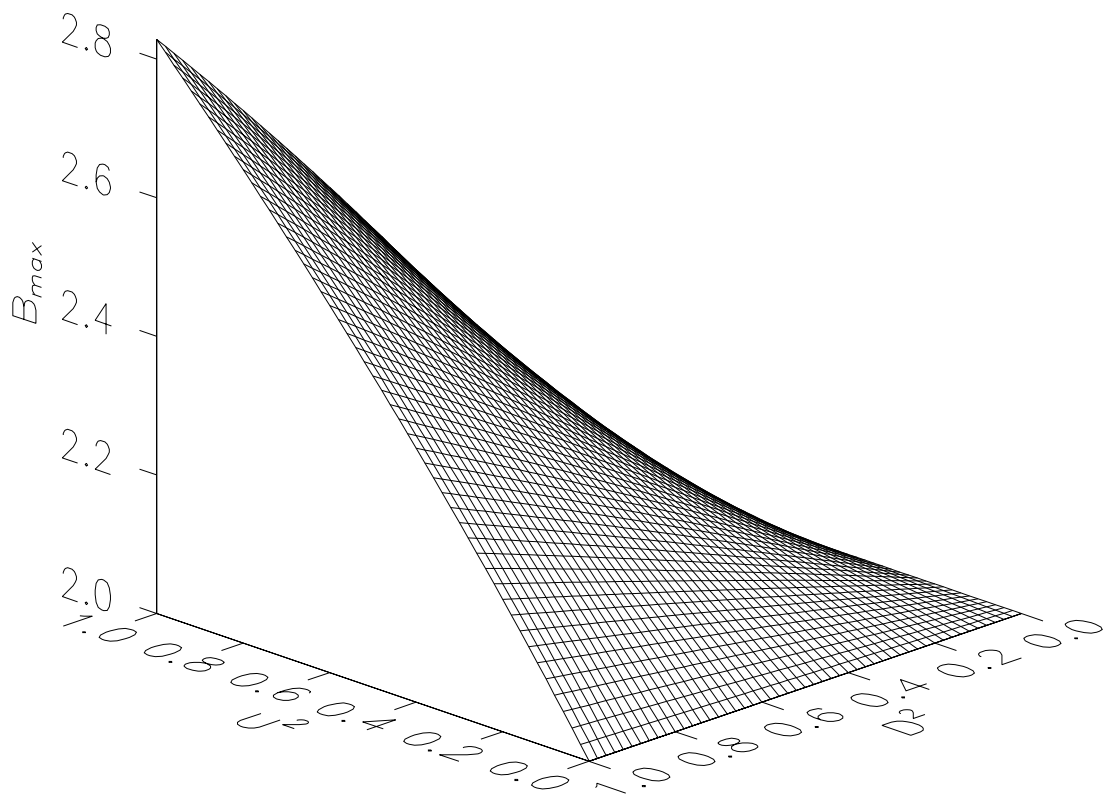
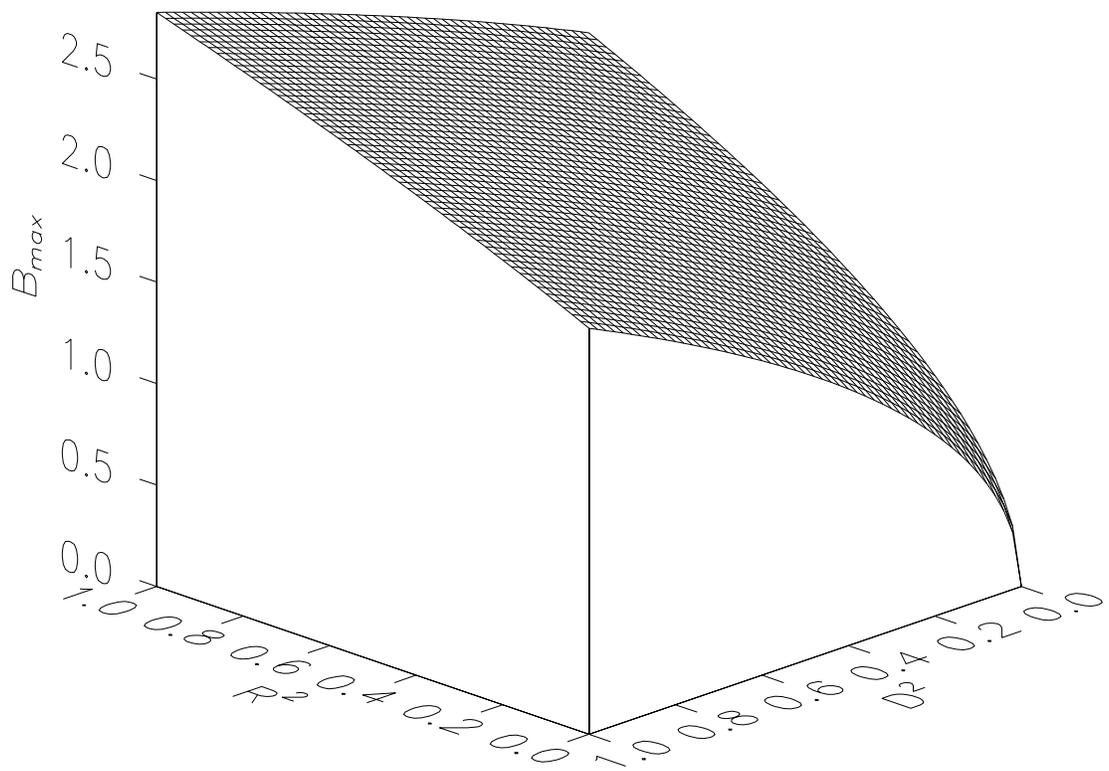
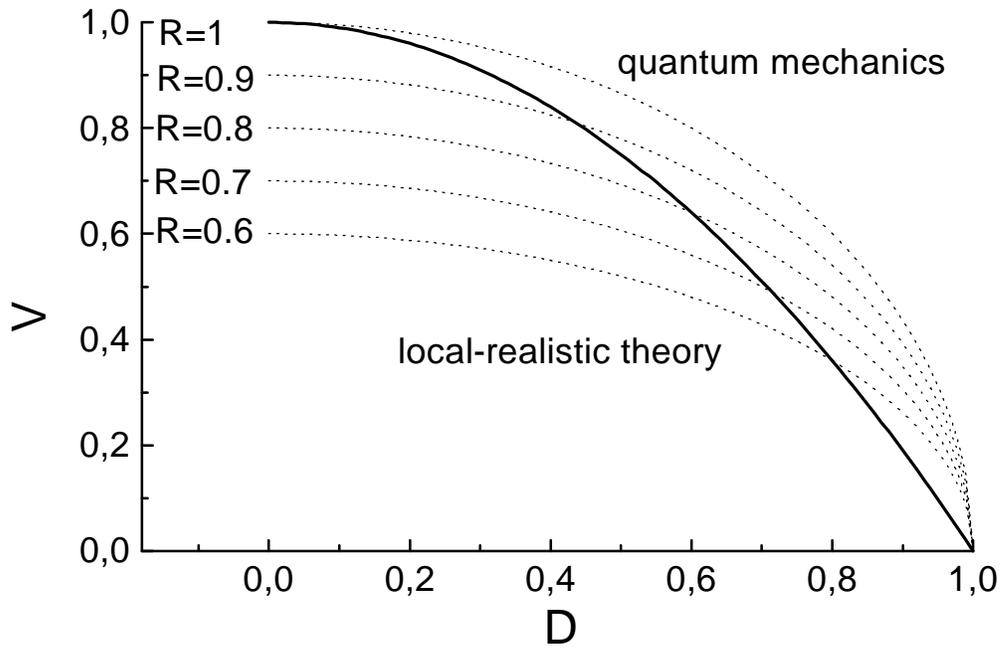
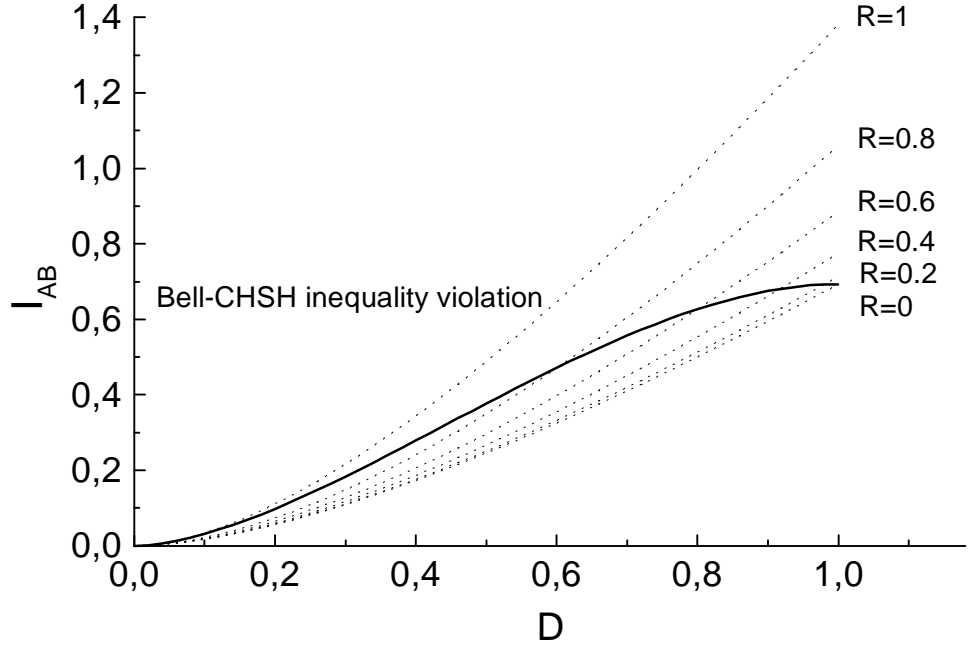
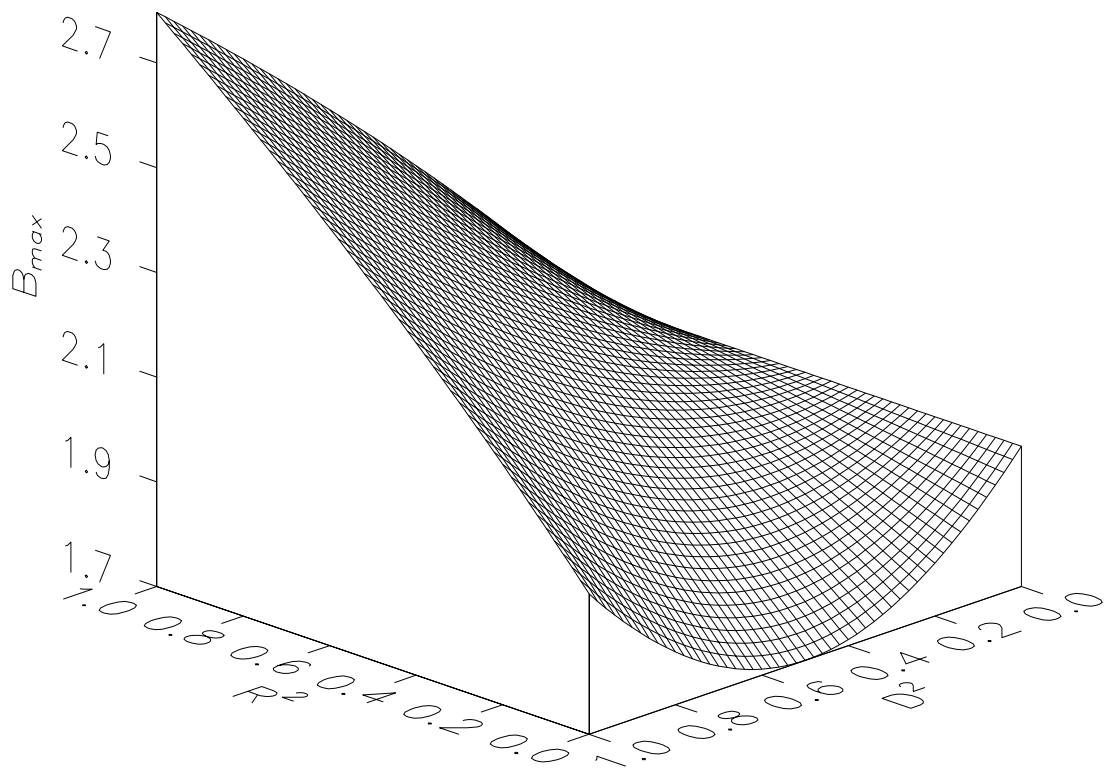
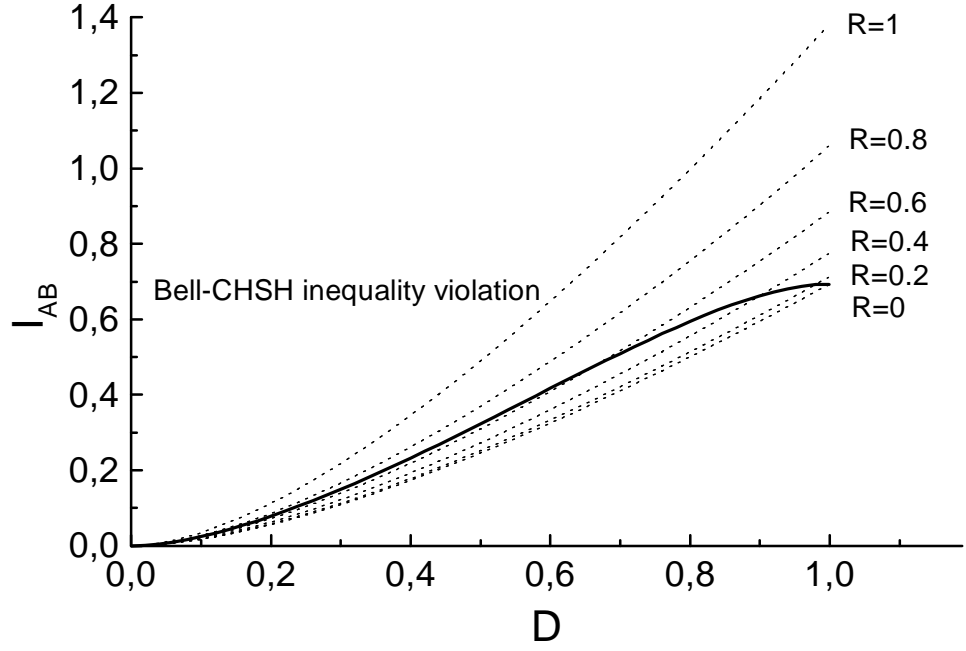
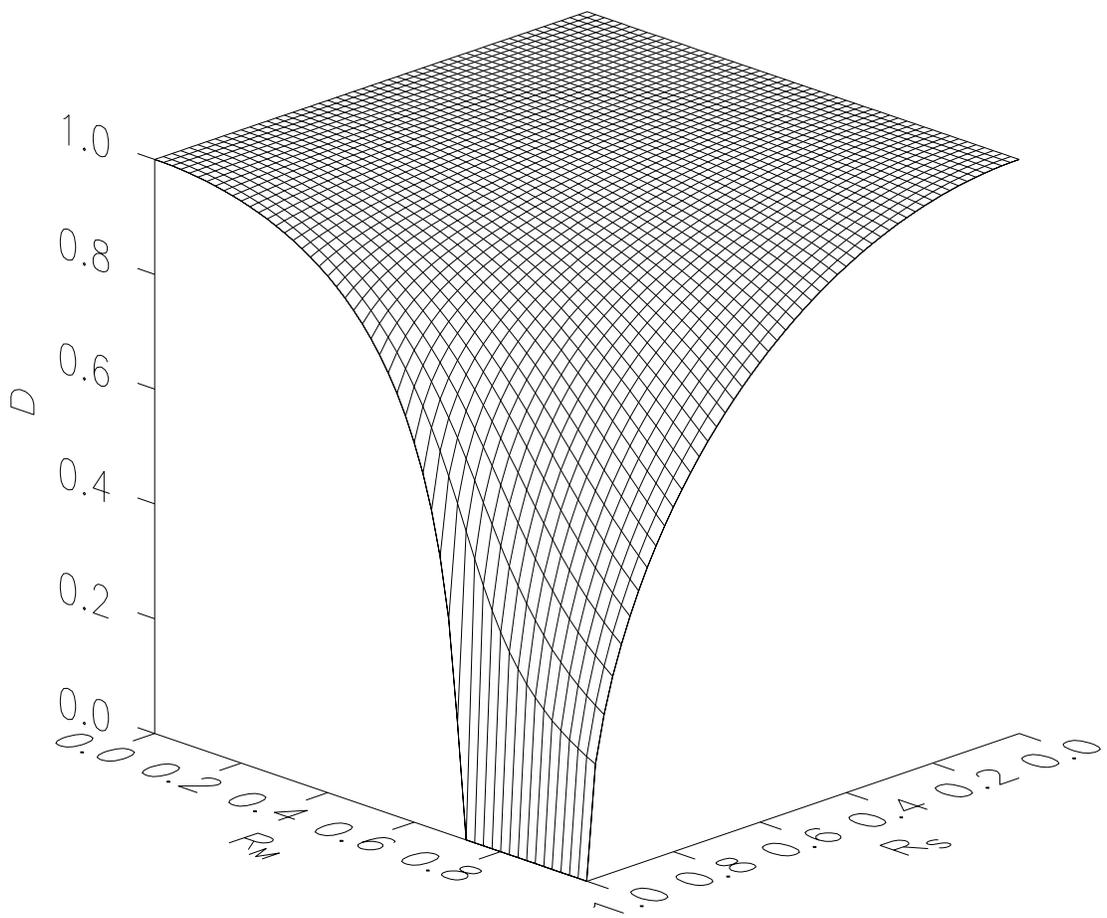
\end{document}